\documentclass[12pt]{iopart}
\usepackage{iopams}  
\usepackage{bm}
\usepackage{epsfig}
\usepackage{graphics}
\usepackage{dsfont}

\def\beq{\begin{equation}}
\def\eeq{\end{equation}}
\def\beqn{\begin{eqnarray}}
\def\eeqn{\end{eqnarray}}

\begin{document}

\title{Dyon condensation in topological Mott insulators}

\author{Gil Young Cho$^{1}$, Cenke Xu$^{2}$, Joel E Moore$^{1, 3}$, and Yong Baek Kim$^{4,5}$}
\address{$^1$ Department of Physics, University of California, Berkeley, CA 94720, USA}
\address{$^2$ Department of Physics, University of California, Santa Barbara, CA 93106, USA}
\address{$^3$ Materials Sciences Division, Lawrence Berkeley National Laboratory, Berkeley, CA 94720, USA}
\address{$^4$ Department of Physics, University of Toronto, Toronto, Ontario M5S 1A7, Canada}
\address{$^5$ School of Physics, Korea Institute for Advanced Study, Seoul 130-722, Korea}
\ead{cgy0417@berkeley.edu, ybkim@physics.utoronto.ca}

\begin{abstract}
We consider quantum phase transitions out of topological Mott insulators in which the ground state of the fractionalized excitations  (fermionic spinons) is topologically non-trivial. The spinons in topological Mott insulators are coupled to an emergent compact U(1) gauge field with a so-called ``axion" term. We study the confinement transitions from the topological Mott insulator to broken symmetry phases, which may occur via the condensation of dyons.  Dyons carry both ``electric" and ``magnetic" charges, and arise naturally in this system because the monopoles of the emergent U(1) gauge theory acquires gauge charge due to the axion term. It is shown that the dyon condensate, in general, induces simultaneous current and bond orders. To demonstrate this, we study the confined phase of the topological Mott insulator on the cubic lattice. When the magnetic transition is driven by dyon condensation, we identify the bond order as  valence bond solid order and the current order as scalar spin chirality order. Hence, the confined phase of the topological Mott insulator is an exotic phase where the scalar spin chirality and the valence bond order coexist and appear via a single transition.  We discuss implications of our results for generic models of topological Mott insulators.
\end{abstract}

\maketitle

\section{Introduction}
One of the most notable progresses in  modern condensed matter physics is the development of the concept of ``topological order''. Topologically ordered phases cannot be understood in terms of order parameters, which are the bases of Landau-Ginzburg theories of broken-symmetry phases. Standard examples of the topological order include quantum Hall phases~\cite{wenrev1, Wen95} and time-reversal symmetric topological  insulators~\cite{fkm,kane&mele2d,mb,roy,hm} (we will abbreviate `time-reversal symmetric topological insulator' as `topological insulator'). Topological  insulators are insulating in bulk but have gapless edge or surface states which are `immune' to the opening up of a gap as long as the time-reversal symmetry is respected. Though the protected gapless edge/surface state itself is quite interesting, it is notable that the topological band insulator can provide a condensed-matter example of axionic electromagnetism~\cite{wilczekaxion,qi1,axion_Essin} 
\beq
L_{\theta} = \frac{\theta}{32\pi^{2}}\varepsilon^{\mu\nu\lambda\rho}F_{\mu\nu}F_{\lambda\rho},
\label{Axion}
\eeq
when the surface is gapped by the time-reversal symmetry breaking perturbations. In condensed matter physics langauge, this axionic term can be understood as a half-integral surface quantum Hall effect~\cite{qi1}. This quantized anomalous Hall effect and other interesting surface and bulk physics of the topological band insulator have attracted both the theoretical and experimental interest~\cite{QiZhang2, JoelExciton,TseMacDonald1,fu1,CM1,Multi2,BiSe,QAH,axion_monopole,Franz_FM,  CM2}.

Axionic electromagnetism, especially in the presence of a dynamical gauge field, has a long history in high-energy physics. The interest in the axion term stems from the $CP$-violation problem~\cite{CP3,CP2,AxionCP} and confinement of quantum chromodynamics~\cite{Hooft}. Though the axion term equation \ref{Axion} is a total derivative, 't Hooft~\cite{Hooft} showed that the axion term is capable of changing the nature of the instantons and the confined phases in non-abelian gauge theories. The phase where the dyon is condensed in the presence of the axion term has a qualitatively different structure from the ordinary confined phase; this difference is traced back to the fact that the monopole receives the gauge charge due to the axion term (this effect is sometimes called ``Witten effect''~\cite{Witten}). Similar behavior can emerge in  abelian gauge theories on the lattice; Cardy~\cite{Cardy2,Cardy1} studied the phase diagram of the gauge theory with the axion term and demonstrated  oblique confinement in the discrete ${\mathbb Z}_{p}$ gauge theory and the compact $U(1)$ gauge theory. The structure of the phase diagram and the related $SL(2;{\mathbb Z})$ symmetry are well studied~\cite{Cardy2,Cardy1,SL2Z}. 

As we now have a condensed matter system exhibiting axion electromagnetism, we are led to the natural question: could we find an example where the axion term changes the nature of the phases and phase transitions in the condensed matter system? 
%As the axion term reveals itself via Witten effect, we can easily anticipate that the confined phase (or condensate of the monopoles) with the axionic electromagnetism will be qualitatively different from an ordinary confined phase. 
The monopole in the presence of the axion term becomes a dyon with the fractional charge~\cite{Hooft, Witten, Wilczek_Dyon}, and the structure of the confined phase should reflect this fractional charge of the dyon. 
In this paper, we will seek to find a condensed matter example of this novel confined phase which reflects the axionic electromagnetism and dyon condensation. To study the confinement, it is important to allow magnetic monopoles in the excitation spectrum. At the same time, the compactness of the $U(1)$ gauge theory is important for the existence of monopoles. Hence, to study the confined phase with Witten effect, we need both a {\it compact} $U(1)$ gauge theory (or non-Abelian gauge theory) and the axion term. Clearly, the usual electronic topological band insulators with the non-compact ``external" electromagnetic fields cannot provide such an example. In the condensed matter system, the compact $U(1)$ gauge theory on the lattice often emerges as a consequence of  ``fractionalization'' of the electron. 
It has been suggested that the so-called topological Mott insulator, where the fractionalized excitations or spinons possess topological band structure, satisfies
%There is a theoretically conjectured phase satisfying 
both requirements, i.e., fractionalization and topological band structure~\cite{TMI1}. 
%, namely the topological Mott insulator~\cite{TMI1}. 

There may be two different routes to obtain topological Mott insulators. First, one could start from an electronic topological band insulator~\cite{TMI1}. Upon increasing the on-site Hubbard interaction, there may be a transition to a spin liquid Mott insulator. Such a transition can conveniently be described by the framework of the slave-rotor field theory, where the electron operator is written as the product of a fermonic spinon operator and a bosonic rotor operator with proper constraint on the Hilbert space. Thus the low energy effective field theory naturally becomes a U(1) gauge theory, where the fermonic spinons and rotor bosons are coupled to a U(1) gauge field.
Here the fermonic spinons carry the spin-1/2 quantum number and the rotor bosons the electric charge of the electron. The transition from the spin liquid to the electronic topological band insulator can be regarded as the bose condensation transition of the rotor bosons. In the spin liquid phase, the charge-carrying rotor bosons are gapped and the fermonic spinons would inherit the topological band structure of the electrons. These fermonic spinons are in turn coupled to an emergent U(1) gauge field as described above.
When the fermionic spinons are integrated out, the non-trivial topology of the spinon spectra leads to an axion term for the emergent U(1) gauge field. In the second case, one may start from spin models~\cite{TMI2} with frustrated antiferromagnetic and ferromagnetic exchange couplings. It was recently shown that slave-fermion theory of such spin models can support a U(1) spin liquid, where the fermonic spinons 
acquire an emergent spin-orbit coupling. In some frustrated lattices, this leads to the topological band structure of the spinons in analogy to the non-interacting electronic topological band insulators. In this case, the non-trivial topology of the spinon spectra is emergent and not inherited from the electrons.  Similarly to the first case, the axion term would arise upon integrating out the spinons in the bulk (we refer the readers to the reference~\cite{TMI1, TMI2} for the detailed discussion of the transition and phases)

In order to gain some insight as to the dyon condensate in the topological Mott insulator, we start by briefly reviewing the confined phase and the Higgs phase~\cite{Motrunich2,Motrunich1}, driven from a ``bosonic" $U(1)$ spin liquid by condensing the monopole and the ``bosonic" spinon. The condensation of bosonic spinons leads to the conventional Neel phase. On the other hand, we obtain the Valence Bond Solid (VBS) state if we condense the monopoles in the $U(1)$ spin liquid in $(3+1)$D, which can be understood as the confinement of electric fluxes. In the VBS state, we have confined spinons and no ``photon". As there are significant connections between the ordinary confined phase and the oblique confined phase, we discuss first the nature of the ordinary confined phases. First, the non-trivial spatial patterns of VBS originate from the crystal momentum carried by the monopole operators. The crystal momentum can be traced back to the nontrivial Berry phase due to the background staggered $U(1)$ gauge charge. The staggered $U(1)$ gauge charges induce the background ``electric" fluxes connecting charges. When a monopole hops around the dual lattice, it sees the ``electric" fluxes and feels the Aharanov-Bohm phase. This is exactly dual to the standard Aharanov-Bohm phase for electric charge circulating around magnetic flux. Due to the Berry phase for the monopole, the monopole tends to condense at a finite momentum. Condensation of a monopole at a finite momentum implies that there is a non-trivial spatial pattern of the monopole current~\cite{Motrunich2,Motrunich1}. Then, the monopole current will induce the ``electric" fluxes just like an electric current inducing a magnetic field. As the valence bond operator is proportional to the ``electric" flux, we end up with patterns of valence bonds.  

Some of the above discussions for the standard confined phase can be applied to the confined phase driven by dyon condensation in the topological Mott insulators. First of all, the dyon will experience a Berry phase if there is a nontrivial background charge pattern because the dyon carries both the ``magnetic" and ``electric" charges. It will be shown in the main text that there exists another contribution to the Berry phase due to the axion term in the topological Mott insulator. Secondly, when the dyon with a non-zero crystal momentum condenses, the system supports both the ``electric" current and ``magnetic" monopole current. The dyon current then will induce the `magnetic" fluxes as well as the ``electric" flux. The ``magnetic" flux should be naturally accompanied by the gauge-charge current in the oblique confined phase. Hence, we obtain a nontrivial phase where the gauge-charge current order {\it coexists} with the bond order when the dyon condenses. 

There are a number of subtle issues in dyon condensation and the physical interpretation of such phenomena. First, it is known that the statistics of the dyons may not be trivial~\cite{stat5,stat4,stat3,stat1,stat2,Goldhaber2,Goldhaber1,Wilczek2,Wilczek1}. We show that there is no statistical transmutation for the dyons in the topological Mott insulator. Thus the dyons are bosons as far as the monopoles are bosonic~\cite{stat1, Wilczek2}. Secondly, we argue that the monopoles in topological Mott insulators have as their quantum numbers only the $U(1)$ gauge charge and crystal momenta.  (From the discussion of the usual topological band insulator~\cite{qi1, axion_Essin}, we physically understand the gauge charge of a monopole as a `polarization' charge so that the dyon is not  a bound `particle' to the monopole. This understanding of the charge of the dyon perhaps implies that there will not be a statistical transmutation as the charge is more like a polarization `cloud', not a particle). The actual pattern of the dyon condensate depend on the details of the microscopic physics, i.e., the lattice structure and the background charges. In this work, we discuss the dyon condensation patterns and their physical implications on the cubic lattice.

The rest of the paper is organized as follows.  In section II, the field theory of dyons on the dual space/lattice is derived starting from the U(1) gauge theory with axion term on the direct space/lattice. We discuss possible dyon condensation patterns and resulting confined phases in section III. We summarize our results and conclude in section IV.

\section{The field theory of dyons on the dual lattice}

We consider fermionic spinons on a cubic lattice for simplicity to develop the basic idea of dyon condensation.  In the following section we connect the general considerations in this section to possible phases proximate to the topological Mott insulator state.
%discussed by Pesin and Balents for certain pyrochlore compounds.

In order to derive the action for the dyons (or monopoles), we first need to integrate out the gapped matter fields in the system and obtain an effective gauge theory. As the fermionic spinons are in the topological insulator state, the topological nature of the spinon states leads to an axion term in the effective $U(1)$ gauge theory as follows:
\beq
L_{A} = L_{QED} + L_{\theta}
\label{Gauge}
\eeq
The first term $L_{QED}$ describes lattice quantum electrodynamics in $(3+1)$D in the presence of the staggered gauge charge 
\beq
L_{QED} = \frac{1}{4g^{2}} (\partial_{\mu}A_{\nu}-\partial_{\nu}A_{\mu})^{2}+ i\sum_{r} A_{\tau}(r) N(-1)^{x+y+z},
\eeq
where $N$ is the magnitude of the staggered gauge charge for the $U(1)$ gauge field residing on the two sublattices of the (bipartite) cubic lattice. That is, there exist background gauge charges $(-1)^{x+y+z} N$ at the lattice points $(x,y,z)$. As will become clear later, these background charges are the source of the Berry phases experienced by monopoles (or dyons). The existence of the background charges on the lattice for the monopoles/dyons can be seen from the following argument. If there were no background charge, the dispersion of monopoles/dyons on the dual lattice would have the minimum at the zero crystal momentum. The confinement transition via the condensation of monopoles/dyons would occur at the zero crystal momentum. This would suggest that the resulting paramagnetic insulator (with a charge gap) would not break any translational symmetry and at the same time there would be no ground state degeneracy. When the number of spins/fermions in the unit cell is odd, this is not possible because the translationally invariant paramagnetic ground states in such cases can only have either a gapped phase with a ground state degeneracy or a gapless phase via the Hasting's theorem~\cite{Hastings}. Thus it is clear that there should be a background charge for the monopoles/dyons in the case of the odd-number of spins per unit cell. When the number of 
spins/fermions per unit cell is even, which is the case in the topological Mott insulator, this is not guaranteed as the Hastings' theorem cannot directly be applied.
Nevertheless, in light of the discussion for the case of the odd-number of spin per unit cell, 
it is conceivable that there should at least exist the situations where there is a non-trivial background charge for the monopoles/dyons. This is the case that we study in our work.

The second term $L_{\theta}$ of equation \ref{Gauge} is the axion term for the emergent compact $U(1)$ gauge field, arising from the topological properties of
the spinons:
\beq
 L_{\theta} =  \frac{\theta}{8\pi^{2}}\varepsilon^{\mu\nu\lambda\rho}\partial_{\mu}A_{\nu}\partial_{\lambda}A_{\rho}.
\eeq
To perform a duality transformation on the corresponding action, 
\beq
\fl
S_{A} = \int d\tau d^{3}{\bf r} \frac{1}{4g^{2}} (\partial_{\mu}A_{\nu}-\partial_{\nu}A_{\mu})^{2}  + \frac{\theta}{8\pi^{2}}\varepsilon^{\mu\nu\lambda\rho}\partial_{\mu}A_{\nu}\partial_{\lambda}A_{\rho}  + i\sum_{r} A_{\tau}(r) N(-1)^{x+y+z}, 
\label{gauge}
\eeq
we introduce the monopole current $m_{\mu} = \frac{1}{4\pi} \varepsilon^{\mu\nu\lambda\rho} \partial_{\nu}S_{\lambda\rho}$, where $S_{\lambda\rho} \in 2\pi {\mathbb Z}$ so that $m_{\mu} \in {\mathbb Z}$. Introduction of the monopole field $m_{\mu}$ modifies $S_{A}$ such that $\partial_{\mu}A_{\nu}- \partial_{\mu}A_{\nu} \rightarrow \partial_{\mu}A_{\nu}- \partial_{\mu}A_{\nu} + S_{\mu\nu}$. Since the action is quadratic in $A_{\mu}$,  we can formally integrate out $A_{\mu}$. In order to do so, we first solve the Gauss law constraint for the ``electric" field of the $U(1)$ gauge theory, namely
$\nabla \cdot {\vec E} = N(-1)^{x+y+z}$, by breaking the gauge field $A_{\mu}$ into the fluctuating part $a_{\mu}$ (which is free of the constraint) and the static field $a^{0}_{\mu}$ that is responsible for the staggered gauge charge. That is, we write $A_{\mu} = a_{\mu} + a^{0}_{\mu}$ and integrate out $a_{\mu}$. This generates only two terms that are simple and intuitively understandable:
\beq
S = \int d\tau d^{3}{\bf r}\quad \left( \frac{G^{2}}{2} m^{\mu}\frac{1}{-\partial^{2}} m_{\mu} + i (X_{0}^{\mu}+\frac{\theta}{2\pi}a^{0}_{\mu})m^{\mu} \right).
\label{result}
\eeq 
The first term describes the Coulomb interaction $\sim - \frac{1}{\partial^{2}}$between dyons with the modified strength $G^{2} = (\frac{g\theta}{2\pi})^{2} + \frac{4\pi^{2}}{g^{2}}$ due to the ``electric" charge $g/2 = g \times \frac{\theta}{2\pi}$ with $\theta= \pi$ and ``magnetic" charge $2\pi$. The second term represents the Berry phase experienced by the dyons due to the staggered QED charge $\nabla \cdot {\vec E} = N(-1)^{x+y+z}$. Here, $X_{0}^{\mu}$ is the dual $U(1)$ gauge field for $a^{0}_{\mu}$. This term has the form of the minimal couplings of $m_{\mu}$ to $a^{0}_{\mu}$ and $X^{0}_{\mu}$, and simply dictates that the monopole carries the $U(1)$ gauge charge $\theta/2\pi$ and the unit magnetic charge $2\pi$ as expected from the axion term.

We now derive the dyon action in equation \ref{result} rigorously, and also show that there is no long-range statistical interaction between the dyons, that is, that the statistics of the monopoles does not experience any statistical transmutation when they are converted to dyons. We begin with the Lagrangian of the action $S_{A}$ in equation \ref{gauge} with $A_{\mu} = a_{\mu} + a^{0}_{\mu}$ where $a^{0}_{\mu}$ represents the static configuration due to the background gauge charge $N(-1)^{x+y+z}$. Splitting the Lagrangian in terms of $a^{0}_{\mu}$ and $a_{\mu}$, and using $f_{\mu\nu} = \partial_{\mu}a_{\nu} - \partial_{\nu}a_{\mu}$ and $f^{0}_{\mu\nu} = \partial_{\mu}a^{0}_{\nu} - \partial_{\nu}a^{0}_{\mu}$, the resulting Lagrangian can be written as
\beq
L = L_{1}[a_{\mu}, S_{\mu\nu}] + L_{2}[a^{0}_{\mu}, S_{\mu\nu}] + L_{3}[a_{\mu}, a^{0}_{\mu}].
\eeq
The Lagrangian for the purely fluctuating part, $L_1$, consists of the standard QED$_{4}$ supplemented by the axion term in the presence of the monopole currents. 
\beq
\eqalign{ L_{1}[a_{\mu},S_{\mu\nu}] =  \frac{1}{4g^{2}} (f_{\mu\nu}+S_{\mu\nu})^{2} + \frac{\theta}{16\pi^{2}}\varepsilon^{\mu\nu\lambda\rho}f_{\mu\nu}S_{\lambda\rho}+ \cr \qquad \qquad \qquad \frac{\theta}{32\pi^{2}}\varepsilon^{\mu\nu\lambda\rho}S_{\mu\nu}S_{\lambda\rho} + \frac{\theta}{32\pi^{2}}\varepsilon^{\mu\nu\lambda\rho}f_{\mu\nu}f_{\lambda\rho}}
\eeq
The other parts of the Lagrangian are given by
\beq
L_{2}[a^{0}_{\mu}, S_{\mu\nu}] =  \frac{1}{4g^{2}} (f^{0}_{\mu\nu})^{2}+ i \eta^{\mu}a^{0}_{\mu}+ \frac{1}{2g^{2}}f^{0}_{\mu\nu}S_{\mu\nu}+ \frac{\theta}{16\pi^{2}}\varepsilon^{\mu\nu\lambda\rho}f^{0}_{\mu\nu}S_{\lambda\rho}, 
\eeq
where $\eta^{\mu} = N(-1)^{x+y+z} \delta^{\mu,\tau}$ encoding the staggered QED charge,
and the mixed contributions
\beq
L_{3} [a_{\mu}, a^{0}_{\mu}] =  \frac{1}{2g^{2}} f^{0}_{\mu\nu}f_{\mu\nu} + \frac{\theta}{16\pi^{2}}\varepsilon^{\mu\nu\lambda\rho}f^{0}_{\mu\nu}f_{\lambda\rho}.
\eeq
Up to this point, there have been no approximations. We now transform each term into a desired form via approximations justified in the limit of the large mass gaps for the monopole and the spinon. We begin by dropping $L_{3} [a_{\mu}, a^{0}_{\mu}]$ entirely using $\partial_{\mu}f_{\mu\nu} \sim 0$ and $\varepsilon^{\mu\nu\lambda\rho} \partial_{\nu} f_{\lambda\rho} =0$. The first condition $\partial_{\mu}f_{\mu\nu} \sim 0$ is satisfied if the gap for ``charged" excitations is large enough  that there is no free current giving rise to a field strength. In the Coulomb phase, every matter field (including monopoles and dyons) is gapped so this approximation is justified. For $L_{2}$, we solve $\frac{1}{4g^{2}} (f^{0}_{\mu\nu})^{2}+ i \eta^{\mu}a^{0}_{\mu}$ first to fix $a^{0}_{\mu}$ and end up with the following Lagrangian  
\beq
\eqalign{
L_{2}[a^{0}_{\mu}, S_{\mu\nu}] =  \frac{1}{2g^{2}}f^{0}_{\mu\nu}S_{\mu\nu}+ \frac{\theta}{16\pi^{2}}\varepsilon^{\mu\nu\lambda\rho}f^{0}_{\mu\nu}S_{\lambda\rho}, \cr \qquad\qquad \quad= X^{0}_{\mu} m_{\mu} + \frac{\theta}{2\pi} a^{0}_{\mu}m_{\mu},}
\eeq
where the equality $f^{0}_{\mu\nu} = \frac{2\pi}{g^{2}}\varepsilon_{\mu\nu\lambda\rho}\partial_{\lambda}X^{0}_{\rho}$ ($X^{0} \sim$ dual $U(1)$ gauge field) and $m_{\mu} = \frac{1}{4\pi} \varepsilon_{\mu\nu\lambda\rho} \partial_{\nu}S_{\lambda\rho}$ are used. Thus, $L_{2}$ contains all the important Berry phase terms for the monopole operator. Now we integrate out $a_{\mu}$ in $L_{1}$ to investigate the Coulomb interaction between monopoles and the statistical interaction. We choose the radiation gauge for the fluctuating $U(1)$ gauge field $a_{\mu}$ (i.e., we use the ``photon" propagator $\frac{g^{2}}{2} \frac{1}{\partial^{2}}$) to obtain 
\begin{eqnarray}
L_{1} =  \frac{1}{2} (\frac{g\theta}{2\pi})^{2}m_{\mu} \frac{1}{\partial^{2}}m_{\mu} + \frac{1}{4g^{2}}(S_{\mu\nu}^{2} + 2 \partial_{\mu}S_{\alpha\mu}\frac{1}{\partial^{2}}\partial_{\nu}S_{\alpha\nu}) \nonumber\\
\qquad+ \frac{\theta}{32\pi^{2}} \varepsilon^{\mu\nu\lambda\rho}S_{\mu\nu}S_{\lambda\rho} + \frac{\theta}{8\pi^{2}} \partial_{\nu}S_{\mu\nu} \frac{1}{\partial^{2}}\varepsilon^{\mu\alpha\beta\rho} \partial_{\alpha}S_{\beta\rho},
\end{eqnarray}
Straightforward calculation reveals that the first line in the above $L_{1}$ becomes the Coulomb interaction~\cite{Cardy1, stat1} between dyons and the second line {\it exactly} vanishes, i.e., $\frac{1}{32\pi} \varepsilon^{\mu\nu\lambda\rho}S_{\mu\nu}S_{\lambda\rho} + \frac{1}{8\pi} \partial_{\nu}S_{\mu\nu} \frac{1}{\partial^{2}}\varepsilon^{\mu\alpha\beta\rho} \partial_{\alpha}S_{\beta\rho} =0 $ up to the boundary terms. We are discussing the bulk transition without boundary, hence this term can be dropped out completely. In fact, this term corresponds to the $\theta$-dependent statistical interaction between the dyons as discussed in A. Goldhaber {\it et.al.}~\cite{stat1} Hence, the dyons do not receive additional statistical interaction among themselves. This suggests that the dyon in the topological Mott insulator is bosonic and can be condensed. 
Using $\theta = \pi$ and collecting the resulting terms in $L_{1}+L_{2}+L_{3}$, we arrive at the effective action for the dyons 
\beq
S = \int d\tau d^{3}{\bf r}\quad \left(\frac{G^{2}}{2} m^{\mu}\frac{1}{-\partial^{2}} m_{\mu} + i (X_{0}^{\mu}+\frac{1}{2}a^{0}_{\mu})m^{\mu}\right),
\eeq
as advertised before. To proceed further, we introduce the soft-spin operator $\psi^{\dagger} \sim \exp(-i\phi)$ for the dyon creation 
and put the lattice structure back to obtain the Hamiltonian on the dual lattice.
\beq
H = - t \sum_{R,R'} \psi^{\dagger}_R \exp(-i(X_{0}^{RR'}+\frac{1}{2}a^{0}_{RR'}+GL_{RR'})) \psi_{R'}+h.c.,
\label{DyonAction}
\eeq
 where we introduce an auxiliary fluctuating $U(1)$ gauge field $L_{RR'}$ to encode the Coulomb interaction with the coupling constant $G = \sqrt{g^{2}/4+4\pi^{2}/g^{2}}$, and $R$ denotes a dual lattice site. 

Some comments on this Hamiltonian are in order. First of all, we need to be careful in defining $a^{0}_{RR'}$ on the dual link as $a^{0}_{\mu}$ was originally meaningful on the links of the direct lattice. To resolve this issue, we need to introduce a short-ranged function $F^{RR'}_{rr'}$ which maps the direct link to the dual link. This defines $a^{0}_{RR'} = \sum_{<rr'>}F^{RR'}_{rr'}a^{0}_{rr'}$. The appearance of this short-ranged function $F^{RR'}_{rr'}$ can be traced back to the form of the axion term which connects $f_{\mu\nu}$ to $f_{\lambda\rho}$ by $\varepsilon^{\mu\nu\lambda\rho}$ and we are forced to introduce such a function~\cite{Cardy2,Cardy1} if we wish to study the axion term on the lattice. Secondly, we have introduced an auxiliary non-compact $U(1)$ gauge field $L_{RR'}$ which differs from $a_{RR'}$ and $X_{RR'}$. In fact, this fluctuating gauge field corresponds to the $U(1)$ gauge field which is {\it rotated} in the electric-magnetic charge. This gauge field is introduced only for convenience as there is no corresponding physical charge.  The two physical charges in the dyon theory are the ``electric" charge and ``magnetic" charge of the compact $U(1)$ gauge theory $A_{\mu}$. However, dyons carry  both charges, and the effective Coulomb interaction between the dyons is the sum of the Coulomb interactions of ``electric" charges and ``magnetic" charges. Because of this, the dyons are seen to interact with the gauge charge $G=\sqrt{g^{2}/4+4\pi^{2}/g^{2}}$. This interaction can be written compactly at the cost of introduction of the auxiliary $U(1)$ gauge field $L_{RR'}$. As the low-energy physics is determinded by the static component, we can ignore this to study the mean-field physics~\cite{Motrunich1}. In the case when $a^{0}_{rr'} = 0$ can be choosen (no flux threading the direct plaquette), we need to solve the following single-particle hopping problem  
\beq
H = - t \sum_{R,R'} \psi^{\dagger}_{R} \exp(-iX_{0}^{RR'}) \psi_{R'}+h.c.
\label{Effective}
\eeq     
This Hamiltonian leads to a set of degenerate energy minima for $\psi_{R}$. This degeneracy will be lifted by the non-linear terms mixing these degenerate minima. This is the same procedure as used in studies of monopole condensation and the resulting confinement problem.

There is, however, an important difference from the monopole condensate problem: the number of dyon species. We now argue that it is enough to consider one doublet of dyon operators in the low energy limit. To see this, we first note that the monopole of the strength $2\pi$ can collect the $U(1)$ charge $n+1/2, n\in {\mathbb Z}$ when the axion angle $\theta = \pi$. Then each dyon will have the self-energy $E \sim (n+1/2)^{2} g^{2} + 4\pi^{2}/g^{2}$. Thus, the dyon of the monopole strength $2\pi$ has two degenerate states with the charge $1/2$ and $-1/2$, representing the lowest energy dyons. We label $\psi_{1}$ and $\psi_2$ as the dyons of the magnetic and electric charge content $(2\pi, 1/2)$ and $(2\pi, -1/2)$. These two fields $(\psi_{1}, \psi_{2})$ are connected to each other by the time reversal symmetry as the time-reversal operation $T$ will flip only the strength of the magnetic field, i.e., $T[\psi_{1}] = \psi_{2}^{\dagger}$ and $T[\psi_{2}] = \psi_{1}^{ \dagger}$ where $\dagger$ operation (particle to anti-particle operation) flips both of the magnetic and electric charge.

The other way to understand the nature of the doublet is to consider the time-reversal symmetry of the spectrum of the dyon fields. We have the time-reversal symmetry before we condense the dyons. As we have a dyon $\psi_{1}$ of $(2\pi, 1/2)$, we need to have a time-reversal partner of this state, i.e., $(-2\pi, 1/2)$ which is nothing but the state $\psi^{\dagger}_{2}$. Hence, we see that when $\psi_{1}$ (or $\psi_{2}$) is condensed, we enter a phase where the time-reversal symmetry is broken.  We note that even though the above derivation is for the $U(1)$ gauge theory, similar calculation for ${\mathbb Z}_{p}$ lattice gauge theory should be possible and indeed has been done by J. Cardy~\cite{Cardy2,Cardy1} without the background charges. 

\section{Dyon condensation and resulting broken symmetry phases}

In this section, we study the broken symmetry phases that arise when the condensation of dyons occurs in the topological Mott insulator. As the precise ordering depends on the lattice structure, we specifically work on the cubic lattice. To determine the patterns of the symmetry breaking explicitly, we work out the lattice symmetry of the dyon fields (more precisely, we determine the projective symmetry group~\cite{r1,r2,r3} of dyon operators) and solve the dyon hopping problem. The effect of  lattice symmetries (at least translations and rotations) on the monopole operator has been thoroughly studied in the literature and we build on those results~\cite{Motrunich1}. We consider the dyon hopping problem on the dual lattice defined as
\beq
H = - t \sum_{a=1,2;R,R'} \psi^{\dagger}_{a,R} \exp(-iX_{0}^{RR'}) \psi_{a, R'}+h.c,
\eeq   
where we {\it choose} $X_{0}^{RR'}$ that is consistent with the background charge $(-1)^{x+y+z}$.
Although here we work on the cubic lattice with the staggered background charge $N = 1$ in equation \ref{Effective}, it should be straightforward to generalize this for different bipartite lattices and larger $N \in {\mathbb Z}$. For example, a larger staggered charge $N=2$ on the cubic lattice has been studied for the ordinary confined phase, and $N=2$ background charges were found to induce similar patterns as the case of $N=1$~\cite{Motrunich2}. 

In the case of $N=1$~\cite{Motrunich1}, we have two minima for each $\psi_{a}$ where the dyon operator takes its minimum kinetic energy. We label these two minima by $\sigma =1, 2$ and hence we have a total of four operators $\psi_{a,\sigma}, a=1,2, \sigma =1,2$. The corresponding eigenfunctions~\cite{Motrunich1} have the forms:
\beq
\eqalign{
\psi_{a,1} = \frac{1+(\sqrt{3}-\sqrt{2})e^{i\pi z}}{\sqrt{2(3-\sqrt{6})}} \cr
\psi_{a,2} = \frac{1-(\sqrt{3}-\sqrt{2})e^{i\pi z}}{\sqrt{2(3-\sqrt{6})}}\times \frac{e^{i\pi x} - ie^{i\pi y}}{\sqrt{2}}.}
\eeq
Hence, we introduce the dyon operators as follows.
\beq
\Psi (R, \tau) = \sum_{a,\sigma} \Psi_{a,\sigma} \psi_{a,\sigma} (R,\tau)
\eeq
The operations of the spatial symmetries on $\Psi_{a,\sigma}$ are worked out in the paper~\cite{Motrunich1}, and quoting their results, we get 
\beq
\eqalign{
T_{x}: (\Psi_{a,1}, \Psi_{a,2}) \rightarrow (\Psi_{a,1}^{*}, -\Psi_{a,2}^{*}) \cr
T_{y}: (\Psi_{a,1}, \Psi_{a,2}) \rightarrow (\Psi_{a,1}^{*}, \Psi_{a,2}^{*}) \cr 
T_{z}: (\Psi_{a,1}, \Psi_{a,2}) \rightarrow (\Psi_{a,2}^{*}, \Psi_{a,1}^{*}) \cr 
R_{z}: (\Psi_{a,1}, \Psi_{a,2}) \rightarrow (e^{i\pi/4}\Psi_{a,1}^{*}, e^{i\pi/4}\Psi_{a,2}^{*}) \cr 
R_{y}: (\Psi_{a,1}, \Psi_{a,2}) \rightarrow (\frac{\Psi_{a,1}^{*}+\Psi_{a,2}^{*}}{\sqrt{2}}, \frac{\Psi_{a,1}^{*}-\Psi_{a,2}^{*}}{\sqrt{2}}).}
\eeq
Note that the spatial symmetries do not mix the dyons with different $U(1)$ gauge charges, i.e., the translations and rotations do not change the index $a = 1,2$. To change $a$, we need to consider the time-reversal operation on $\Psi_{a,\sigma}$ because $T[\Psi_{1,\sigma}] = \Psi_{2,\sigma}^{*}$ and $T[\Psi_{2,\sigma}] = \Psi_{1,\sigma}^{*}$. The time-reversal operation does not change the index $\sigma$ because (1) the dyon operator in the dual theory does not change the sign (i.e., $T: t_{RR'} \rightarrow t_{RR'}$), and hence no change is involved for the minima of the kinetic energy of the dyon operators (2) the minima of the kinetic energy of the dyon operators are at the time-reversal symmetric momentum. Hence, the spatial symmetry is completely determined by the actions on $\sigma$. We are now ready to introduce the following two vector order parameters 
\begin{eqnarray}
{\vec E}_{r}  = \frac{\Psi_{1,\alpha}^{\dagger} {\vec \sigma}^{\alpha\beta}\Psi_{1,\beta} + \Psi_{2,\alpha}^{T}{\vec \sigma}^{\alpha\beta}\Psi^{*}_{2,\beta}}{2} \nonumber\\
{\vec B}_{r}  = \frac{\Psi_{1,\alpha}^{\dagger} {\vec \sigma}^{\alpha\beta}\Psi_{1,\beta} - \Psi_{2,\alpha}^{T}{\vec \sigma}^{\alpha\beta}\Psi^{*}_{2,\beta}}{2}, 
\end{eqnarray} 
for $r= x,y,z$. For ${\vec V} = {\vec E}, {\vec B}$, we have the following transformation rules 
\begin{eqnarray}
T_{x}: (V_{x}, V_{y}, V_{z}) \rightarrow (-V_{x}, V_{y}, V_{z}) \nonumber \\ 
T_{y}: (V_{x}, V_{y}, V_{z}) \rightarrow (V_{x}, -V_{y}, V_{z}) \nonumber \\ 
T_{z}: (V_{x}, V_{y}, V_{z}) \rightarrow (V_{x}, V_{y}, -V_{z}) \nonumber \\ 
R_{z}: (V_{x}, V_{y}, V_{z}) \rightarrow (V_{y}, V_{x}, V_{z}) \nonumber \\ 
R_{y}: (V_{x}, V_{y}, V_{z}) \rightarrow (V_{z}, V_{y}, V_{x}).
\end{eqnarray}
The vectors ${\vec V} = {\vec E}, {\vec B}$ follow the same spatial symmetry transformation rules because the spatial symmetry is fully determined by the actions on $\sigma$ in $\Psi_{\alpha, \sigma}, \alpha = 1,2, \sigma = 1,2$. In addition to the lattice symmetry, we also need to introduce the time reversal symmetry operations: 
\beq
T: ({\vec E}, {\vec B}) = ({\vec E}, -{\vec B}).
\eeq
In the end, we have two vector operators ${\vec E}$ and ${\vec B}$. What phases do these two vector operators ${\vec E}$ and ${\vec B}$ represent? It can be shown that the field ${\vec E}$ behaves like staggered bond order (equivalently, electric flux) and ${\vec B}$ behaves like  staggered current order (magnetic flux). 

If the confinement induces ordering of the magnetic degrees of freedom, then the dual operators ${\vec E}$ and ${\vec B}$ should have  representations in terms of the spin operators ${\vec S}$~\cite{TMI2}. In the previous studies, it is already noticed that ${\vec E}$ represents the VBS-like order parameters~\cite{Motrunich2,Motrunich1}. More explicitly, $E_{x}  \sim (-1)^{x} <{\vec S}_{{\bf r}}\cdot {\vec S}_{{\bf r}+{\hat x}}>$ and there are analogous identifications for $E_{y}$ and $E_{z}$. This VBS corresponds to ``confined electric flux''.  

It is not difficult to see that the magnetic flux expected in the dyon-condensed phase is the field ${\vec B}$. This magnetic flux should be sourced by the spinon current living on the plane perpendicular to the ${\vec B}$. In turn, the spinon current can be related to the scalar spin chirality. These facts, hand-in-hand, lead us to the conclusion that the scalar spin chirality can be connected to the vector field ${\vec B}$. This intuition is supported by explicit construction of appropriate combinations of the scalar spin chirality, which follow the same transformation rules as ${\vec B}$. 
\beq
B_{x} = (-1)^{x} M_{yz}, B_{y} = (-1)^{y} M_{zx}, B_{z} = (-1)^{z} M_{xy}
\eeq
where $M_{ij}$ is the scalar spin chirality defined on the $ij$-plane. For example, $M_{xy}$ involves four spins of the plaquette on the direct lattice. We label four sites $1 = (x,y,z)$, $2=(x+1,y,z)$, $3 = (x+1,y+1,z)$ and $4=(x,y+1,z)$ in the $xy$-plane. Then, $M_{xy}$ is defined uniformly on every plaquette of the $xy$-plane as 
\beq
\fl
M_{xy} = <{\vec S}_{1}\cdot {\vec S}_{2}\times {\vec S}_{3}> +<{\vec S}_{2}\cdot {\vec S}_{3}\times {\vec S}_{4} >
+<{\vec S}_{3}\cdot {\vec S}_{4}\times {\vec S}_{1}> + <{\vec S}_{4}\cdot {\vec S}_{1}\times {\vec S}_{2}>.
\eeq
It can be shown that $(-1)^{z} M_{xy}$ and $B_{z}$ follow the same transformation rules under the lattice symmetries and time-reversal symmetry operations. Analogous expressions hold for $B_{y}$ and $B_{z}$ (as noted before, the scalar spin chirality lies in the plane perpendicular to the vector ${\vec B}$). Thus, we confirm that the scalar spin chirality is induced in the dyon condensed phase. Now we discuss where the scalar spin chirality order comes from. The current order ${\vec B}$ originates from the `mixing' of the two dyon species with different $U(1)$ gauge charges. This order parameter would have not been present if there were only one monopole species as in the usual confinement problem. In the case of the dyon condensate, both of the electric flux and magnetic flux should be confined at the same time. This implies that there should be sources of the magnetic flux as well as the electric flux. We can easily anticipate that the magnetic flux ($\sim {\vec B}$) can arise from the spinon current ($\sim M_{ij}$) via the elementary Biot-Savart Law.

With these identifications between the dyon operators and the corresponding order parameters, we can now ask what kind of phases will eventually appear. This requires the studies of the quartic order and higher order terms in the Landau-Ginzburg theory~\cite{Motrunich2,Motrunich1} written in terms of the dyon fields ${\vec \Psi}^{T}_{a} = (\Psi_{a,1},\Psi_{a,2}), a=1,2$. We will not attempt to solve the full phase diagram but will point out a few representative confined phases from the symmetry-allowed forms of the Landau-Ginzburg theory 
\beq
L = L_{kin}({\vec \Psi}_{1}) + L_{kin}({\vec \Psi}_{2}) + L_{int} ({\vec \Psi}_{1},{\vec \Psi}_{2} ).
\eeq
Here, $L_{kin}$ represents the kinetic terms in equation \ref{Effective}. We expand $L_{int} = L_{2}+L_{4}+ \cdots$ with $L_{n} \sim O(\Psi^{n}), n\in {\mathbb Z}$. By requiring the lattice symmetry and the time-reversal symmetry, we identify $L_{2} = r |{\vec \Psi}_{1}|^{2} + r  |{\vec \Psi}_{2}|^{2}$ and 
\beq
L_{4} = u({\vec \Psi}_{1}^{\dagger}\cdot {\vec \Psi_{1}})^{2} + u({\vec \Psi}_{2}^{\dagger}\cdot {\vec \Psi_{2}})^{2}+v |{\vec \Psi}_{1}|^{2}|{\vec \Psi}_{2}|^{2} +w |{\vec \Psi}_{1}\cdot {\vec \Psi}_{2}|^{2} .
\eeq
Up to this quartic order, we notice that there are several interesting possibilities. First, $<{\vec \Psi_{1}}>= 0$ and $<{\vec \Psi_{2}}>=0$ for $r>0$: This phase is the topological Mott insulator phase. The order parameters are ${\vec E} = 0$ and ${\vec B} = 0$. Secondly, when the sign of the mass term for ${\vec \Psi_{a}}$ is negative $r<0$ in $L_{2}$, two dyon fields ${\vec \Psi}_{1}$ and ${\vec \Psi}_{2}$ tend to condense. We assume $u>0$ in $L_{4}$. There are four following possibilities of $L_{4}$. \newline${}$\newline$({\bf 1})$ If $w<0$ and $2u>v>-2u$, we have $<{\vec \Psi_{1}}>={\vec \rho}\exp(i\phi_{1}), <{\vec \Psi_{2}}>= {\vec \rho}\exp(i\phi_{2})$ where ${\vec \rho}$ is a real two-component spinor. Then, the order parameter ${\vec E} \propto {\vec \rho}^{T}\sigma {\vec \rho} \neq 0$ but ${\vec B} = 0$. The two dyon fields condense at the equal amplitude in a way to respect the time-reversal symmetry. Hence ${\vec B} =0$ is inevitable, and this phase is the ordinary confined phase. \newline${}$\newline$({\bf 2})$ If $w>0$ and $v>2u$, then we have $<{\vec \Psi}_{1}> \neq 0$ but $<{\vec \Psi_{2}}>=0$. Here, ${\vec E}$ and ${\vec B}$ acquire non-zero expectation values (with ${\vec E} \parallel {\vec B}$) at the same time, and in this phase, the scalar spin chirality and the VBS order coexist. Furthermore, the time-reversal symmetry is spontaneously broken. \newline${}$\newline$({\bf 3})$ If $w>0$ and $2u>v+w>-2u$, we have $<{\vec \Psi_{1}}>={\vec \rho}_{1}\exp(i\phi_{1}), <{\vec \Psi_{2}}>= {\vec \rho}_{2}\exp(i\phi_{2})$ where $\rho_{1}$ and $\rho_{2}$ are the real two-component spinors with ${\vec \rho}_{1}\cdot {\vec \rho}_{2}=0$. In this confined phase, the order parameters ${\vec E}=0$ but ${\vec B} \neq 0$. Thus, this phase corresponds to the phase where only the scalar spin chirality orders. Here, the two dyon fields condense at the equal amplitude in a way to break the time-reversal symmetry {\it maximally}. This phase should be considered as an ``opposite'' of the phase $({\bf 1 })$ where only ${\vec E}$ acquires non-zero expectation value. \newline${}$\newline$({\bf 4})$ If $w<0$ and $v+w>2u$, the only one of the two dyon fields condenses. The phase is identical to the phase $({\bf 2})$ and there are two orders present in the phase. \newline${}$

The detailed spatial patterns of the order parameters can depend on the higher-order terms in $L_{int}$ which we do not consider (in fact, the pattern depends on the $\sim O(\chi^{8})$ in the Landau-Ginzburg theory, see for example Motrunich and Senthil~\cite{Motrunich1}).

We would like to point out that VBS order can be detected by measuring the triplon spectrum in inelastic-neutron scattering~\cite{V2} combined with the magnetic susceptibilities measurements~\cite{V1}.  The scalar spin-chirality may be detected via Raman scattering~\cite{R1,R2} or resonant inelastic x-ray scattering(RIXS)~\cite{R3}. 

\section{Conclusion and outlook}
In this paper, we have studied the condensation of dyons and the corresponding confined phases in topological Mott insulators. Using the effective field theory and projective symmetry analysis on the dyon operators, we identified possible broken symmetry phases resulting from such confinement. Because the dyons are monopoles endowed with gauge charge, the confined phase can induce at least two different orders:  bond order and current order. In the case when the axion angle $\theta$ is not $\pi$, we might need to consider more species of the dyons. However, the dyon carries the ``electric'' and ``magnetic'' charges as far as $\theta \neq 0$, and this implies that the confined phase in the non-zero axion angle $\theta$ will be a coexisting phase of the bond order and the current order if $\theta \neq 0$. 
In this paper, we concentrate on possible confined phases on the cubic lattice. but our results should be straightforwardly generalizable to other bipartite lattices, 
where the precise spatial patterns of bond and current orders would depend on the lattice structure.

For the physical spin/magnetic degree of freedom, the physical order parameters in the confined phase were shown to be the scalar spin chirality order and the VBS order. It should be emphasized that the coexistence of two order parameters discussed above is unique to the confined phase of the topological Mott insulators and does not arise in ordinary confinement physics in condensed matter systems (monopoles are not dressed with any quantum numbers in the ordinary confined phase~\cite{Motrunich2, Motrunich1}). Notice also that the coexistence of the VBS and scalar spin chirality order is difficult to obtain in a Ginzburg-Landau theory framework in ordinary condensed matter systems as this would require extremely high-order coupling between spin operators while this can occur naturally if one starts from topological Mott insulators. Thus the emergence of the coexisting scalar spin chirality and VBS orders via a second order transition from a quantum paramagnet would be strong evidence that such a quantum paramagnet was in fact the topological Mott insulator. 

\ack
We thank S. Bhattacharjee, E.G. Moon, T. Grover, J. Maciejko, S.B Lee, A. Vishwanath and P. Yi for the helpful discussions. G.Y.C especially thank M. Fisher for the insightful discussion on the dyon. This work was supported by KITP graduate fellowship (GYC), the Sloan Foundation (CX), NSF DMR-0804413 (GYC and JEM), the NSERC of Canada and CIFAR (YBK).

\end{document}